\documentclass[11pt,english]{extarticle}
\usepackage[latin9]{inputenc}
\usepackage{bm}
\usepackage{amsmath}
\usepackage{xcolor}
\usepackage{amsthm}
\usepackage{amssymb}
\usepackage[a4paper]{geometry}
\makeatletter
\newcommand{\lyxaddress}[1]{
	\par {\raggedright #1
	\vspace{1.4em}
	\noindent\par}
}

\usepackage{babel}

\makeatother

\providecommand\theoremname{Theorem}
\theoremstyle{plain}
\newtheorem{thm}{\protect\theoremname}
\providecommand\remarkname{Remark}
\theoremstyle{remark}
\newtheorem{rem}{\protect\remarkname}
\providecommand\lemmaname{Lemma}
\theoremstyle{plain}
\newtheorem{lem}{\protect\lemmaname}
\usepackage{babel}
\begin{document}
\title{Coalescing and annihilating random walks on the ring}
\author{A.M. Povolotsky$^{1,2}$, R. Tribe$^{3}$,O.V. Zaboronski$^{3}$ }
\maketitle

\lyxaddress{$^{1}$Bogoliubov Laboratory of Theoretical Physics, Joint Institute
for Nuclear Research, 6 Joliot-Curie, Dubna, Russia\\
$^{2}$National Research University Higher School of Economics, 20
Myasnitskaya, Moscow, Russia\\
$^{3}$Mathematics Institute, University of Warwick, Coventry, CV4
7AL, UK\\
}
\abstract
We study correlation functions in the model of coalescing and annihilating random walks on the one-dimensional periodic lattice $\mathbb{Z}_N$.  
We prove that, for a special class of initial conditions, the $n$-point correlation function can be represented as a linear combination of two Pfaffians of $2n\times 2n$ matrices constructed from $2\times2$ matrix kernels, which in turn can be expressed in terms of two scalar kernels.
This is an analogue of the single-Pfaffian form exhibited by correlation functions of the same model on $\mathbb{Z}$, where it serves as a signature of a Pfaffian point process. 
We derive equations for the kernels and construct their solutions. 
We also show that a particular example of an initial condition leading to bi-Pfaffian correlation functions is the inhomogeneous Bernoulli point process, which includes a fixed initial configuration as a particular limiting case.

\section{Introduction}

Interacting particle systems have attracted significant attention from both physical and mathematical communities. For the former, they
serve as a testing ground for studies in non-equilibrium statistical
physics, and for the latter, they are considered an example of mutually
interacting Markov processes. An important class of interacting particle
systems is that of the reaction-diffusion (RD) systems. In addition
to diffusive motion in space, particles in RD systems are assumed
to disappear from or enter the system as a result of local inter-particle
interactions, which are often interpreted as chemical reactions. Among
the simplest models of this type are annihilating random walks (ARW)
and coalescing random walks (CRW) as well as their continuous analogues,
annihilating and coalescing Brownian motions abbreviated ABM and CBM,
respectively. In these models, particles diffusing in space undergo
reactions of annihilation, $A+A\to\emptyset$, and coalescence, $A+A\to A$,
every time a pair of them meets. 

Initially, interest in these models was motivated by studies in
chemical kinetics. The first results date back to Marian Smoluchowski
\cite{smoluchowski1916drei}, who predicted an inverse linear time
decay of particle density, $\rho(t)=O\left(t^{-1}\right),t\to\infty$,
in systems of pair-annihilating particles within what is now known
as the mean-field approximation. However, in 1976 Ovchinnikov and
Zeldovich \cite{ovchinnikov1978role} pointed out the importance of
fluctuations due to diffusion, which change this law in low dimensions.
Further studies were aimed at accounting precisely for these effects
in various settings. Around the same time, extensive studies of ARW
and CRW in $\mathbb{Z}^{d}$ and ABM and CBM in $\mathbb{R}^{d}$
were also initiated in the probabilistic community (see, e.g., \cite{erdos1974some,griffeath1978annihilating,griffeath1979additive,arratia1981limiting}).
As a result, the mean-field estimates for the particle density were
corrected to $\rho(t)=O\left(t^{-1/2}\right)$ for $d=1$ and to $\rho(t)=O\left(\ln t/t\right)$ for $d=2$, while for $d>2$ the mean-field prediction remained valid
\cite{bramson1980asymptotics}. 

Plenty of results on the models have been obtained since then using both
physical insights and mathematically rigorous arguments. An important
place is taken by the one-dimensional systems as they turned out to
be exactly solvable. In practice, this means that there is a range
of methods applicable (at least potentially) to the derivation of the
complete set of correlation functions of the model, which fully characterize
the law of the process. Among the methods applied to the mentioned
models are mappings to spin chains and free fermions \cite{lushnikov1986binary,schutz2001exactly},
the empty interval method and the method of Markov dualities relating the ARW and CRW models to the voter model or the zero-temperature kinetic Ising model\cite{griffeath1979additive,glauber1963time,felderhof1971spin},
which in turn are dual to ARW. In particular, using the empty interval
method \cite{ben1995method}, the whole set of single-time correlation functions for CBM and ABM was obtained in \cite{ben1998complete}
and \cite{masser2001correlation}, respectively, on the physical level
of rigor. A mathematically rigorous approach to the problem was employed
in \cite{tribe2011pfaffian} with the use of the Markov duality method.
The fixed-time laws of particle configurations in both CBM and ABM
under the maximal entrance law were shown to be Pfaffian point processes,
which under proper rescaling are equivalent to the one characterizing
the real eigenvalues of the real Ginibre ensemble of random matrices \cite{ginibre1965statistical,borodin2009ginibre}.
Furthermore, the results of \cite{tribe2012one,tribe2015erratum}
on multi-time correlation functions established that the whole law for ABM is given by an extended Pfaffian process. In \cite{garrod2018examples}
these results were extended to the model of coalescing and annihilating
random walks (CARW) on $\mathbb{Z}$ in the cases of arbitrary deterministic
and Bernoulli initial conditions. The latter model, which will be
the subject of the present article, unifies the rules of ARW and CRW
by requiring that either annihilation or coagulation happen with complementary
probabilities when two particles meet. 

A natural question in the context of the above studies is how the confined
geometry affects the behaviour of RD systems. In particular, on a finite ring the particle density approaches a stationary
limit exponentially, rather than displaying algebraic long-time relaxation. However,
a more detailed analysis, which in particular could clarify the finite-size scaling form of the correlation functions, is of interest. Though
many results were obtained for the simplest case of the periodic system,
see e.g. \cite{torney1983diffusion,balding1988diffusion,balding1989diffusion1,schutz1996nonequilibrium},
the full characterization of the probability law of the process by
the full set of explicitly given correlation functions is still lacking.
Below, we make a step in this direction. We consider the CARW model on
the finite ring $\mathbb{Z}_{N}$ and derive all the single-time correlation
functions. Unlike the correlation functions of CARW on $\mathbb{Z}$,
which have the Pfaffian form specific to the Pfaffian point process,
the ones on the ring are shown to be given by a linear combination
of two Pfaffians. To our knowledge, no such point process has been
considered yet. Hence, we refer to it as a bi-Pfaffian point process.
The approach we use to this end is the Markov duality method in its
algebraic interpretation presented recently in \cite{povolotsky2025representations}.
Similar results can also be obtained for the continuous limit of the
model, coalescing and annihilating Brownian motions, though the work
with a continuous state space requires an additional analytic effort.
The result on this model will be presented in a separate forthcoming
article \cite{povolotsky2026coalescing}. 

The article is organized as follows. In Section 2 we formulate the
model and state the main results of the article. These are Theorem
\ref{thm: bi-PPP} on the bi-Pfaffian form of the correlation functions
for a special class of initial conditions and Lemma \ref{lem: Inhomogeneouse Bernoulli IC}
that states that the inhomogeneous Bernoulli point process, including its degenerate case of a fixed initial configuration, is in this class.
At the end of this section we briefly discuss the open problems. In
Section 3 we prove Theorem \ref{thm: bi-PPP}. To this end, we first
briefly introduce an algebraic approach to Markov dualities, developed
in \cite{povolotsky2025representations}, which is used to derive
the evolution equations for the duality functions. Then, we show that
the equations for the $2n$-point duality functions are solved in
terms of the Pfaffians of the two-point ones. Then the statement of
the main theorem for the correlation functions is reduced to simple
differentiation of the result obtained. In Section 4 we prove Lemma
\ref{lem: Inhomogeneouse Bernoulli IC} on the suitable choice of
initial conditions, reducing the problem to a simple Pfaffian identity.
Section 5 is devoted to the explicit solution of the equation for
the so-called scalar kernels, which are the main ingredient of the
matrix kernel used in the resulting expression of the correlation
functions.

\section*{Acknowledgments}

AP is grateful to the organizers of the conference ISQT-2026, where the
results of this article were first presented, for their warm hospitality. 

\section{Model, results and discussion}

Below we define the CARW model as a continuous-time Markov process on
the state space $\Omega\cong\{0,1\}^{\mathcal{L}}$. Within this article
$\mathcal{L}$ is assumed to be a one-dimensional periodic lattice
with $N$ sites, $\mathcal{L}=\mathbb{Z}_{N}$ ($N\geq2$). Then, $\Omega$ is
the set of particle configurations on the lattice, which can be recorded
by a string $\eta=\left(\eta(1),\dots,\eta(N)\right)$ of zeroes and
ones, in which the state of site $i$ is either $\eta(i)=1$, when
the site is occupied by a particle, or $\eta(i)=0$, when it is empty.
Below we will also mention the CARW model from \cite{garrod2018examples}
on the 1D infinite integer lattice $\mathcal{L}=\mathbb{Z}$, in which
particle configurations are represented by infinite strings of zeroes
and ones. 

The dynamics of the process is defined as follows. Every particle
is supplied with an independent exponential alarm clock, which rings
after a random waiting time $T$ distributed as $\mathbb{P}(T<a)=1-e^{-a}.$
When the clock has rung, the particle jumps either left or right with
probabilities $0\leq l\leq1$ and $r=1-l$, respectively. If, as a result,
the particle has arrived at an occupied site, the two particles either
instantaneously annihilate, i.e. the site becomes empty, with probability
$0\leq\theta\leq1$, or instantaneously coalesce, i.e. the site becomes
occupied by one particle, with probability $(1-\theta).$ 

Alternatively, one can think of the change of state of a pair of neighboring sites, which happens independently for every pair with rates shown
below. 
\begin{equation}
10\xrightarrow{r}01,\,\,01\xrightarrow{l}10,\,\,11\xrightarrow{r(1-\theta)}01,\,\,11\xrightarrow{l(1-\theta)}10,\,\,11\xrightarrow{\theta}00\label{eq: rates}
\end{equation}

Given a probability distribution $\mu_{0}$ on the state space $\Omega$,
we define the random process $\eta_{\cdot}:\mathbb{R}_{\geq0}\to\Omega$,
$t\mapsto\eta_{t}$, to be a one-parameter family of random particle
configurations $\{\eta_{t}\}_{t\geq0}$ that, being conditioned on
the initial distribution $\mu_{0}(\eta)=\mathbb{P}(\eta_{0}=\eta)$,
evolves according to the specified rules.

The distribution of a fixed-time random configuration $\eta_{t}$
is an example of a simple point process on the set $\mathcal{L}$. For finite or countable $\mathcal{L}$, this is to say that it is a probability
measure on the set $2^{\mathcal{L}}$ of subsets of $\mathcal{L}$.
In the context of the present article, the random subsets are particle
configurations.

A simple point process can be fully characterized by the set of correlation
functions 
\[
\rho_{n}(i_{1},\dots,i_{n};t):=\mathbb{E}\eta_{t}(i_{1})\cdots\eta_{t}(i_{n})=\mathbb{P}(\{i_{1},\dots,i_{n}\}\subset\eta_{t}),\quad i_{1}\neq\dots\neq i_{n}\in\mathcal{L},n=1,\dots,|\mathcal{L}|.
\]
It is known \cite{garrod2018examples} that when $\mathcal{L}=\mathbb{Z}$,
the fixed-time configuration of CARW is a Pfaffian point process (PPP)
for a range of initial conditions. This is to say that there exists
a skew-symmetric matrix function $\mathbf{K}:(\mathcal{L\times}\{1,2\})\times(\mathcal{L\times}\{1,2\})\to\mathbb{R},$
such that all the correlation functions are given by Pfaffians of
its $2n\times2n$ diagonal blocks,
\[
\rho_{n}(i_{1},\dots,i_{n})=\mathrm{Pf}\mathbf{K}|_{I\times I}
\]
for any $n\in1,\dots,|\mathcal{L}|$ and any $I=\{i_{1},\dots,i_{n}\}\subset\mathcal{L}$.
Here, the r.h.s. should be understood as the Pfaffian of a $2n\times2n$ skew-symmetric block matrix with $2\times2$ blocks 
\begin{equation}
\mathbf{K}(i,j)=\left(\begin{array}{cc}
\mathbf{K}_{11}(i,j) & \mathbf{K}_{12}(i,j)\\
\mathbf{K}_{21}(i,j) & \mathbf{K}_{22}(i,j)
\end{array}\right),\label{eq: 2 x 2 blocks}
\end{equation}
and the Pfaffian of a skew-symmetric $2n\times2n$ matrix $A$ is defined
as 
\begin{equation}
\mathrm{Pf}A\equiv\mathrm{Pf}(A_{ij};1\leq i<j\leq 2n)=\sum_{\pi\in P_{2}([2n])}\mathrm{(-1)^{|\pi|}}A_{\pi(1)\pi(2)}\cdots A_{\pi(2n-1)\pi(2n)},\label{eq: Pfaffian def}
\end{equation}
where the sum is over the set $P_{2}([2n])$ of pairings $\pi=\{\{\pi(1),\pi(2)\},\dots,\{\pi(2n-1),\pi(2n)\}\}$
of elements of the set $[2n]=\{1,\dots,2n\}$ and $|\pi|$ is the
number of crossings within the pairing. The second notation defines
the Pfaffian in terms of the explicit coefficients of the upper-triangular
part of $A$. 

Now we formulate the main result of the present article, which is
a direct generalization of the result of \cite{garrod2018examples}
for the case of the ring. 
\begin{thm}
\label{thm: bi-PPP}Let $\{\eta_{t}\}_{t\in\mathbb{R}_{+}}$ be CARW
on $\mathcal{L}=\mathbb{Z}_{N}$ conditioned on an initial distribution
$\mu_{0}$. In particular, we define 
\[
\Phi:=\mathbb{E}\left(-\theta\right)^{\left|\eta_{0}\right|},
\]
where $\left|\eta_{0}\right|=\sum_{i\in\mathcal{L}}\eta_{0}(i)$.
Let us assume that $\Phi\neq\pm1.$ Then, for any $1\leq n\leq N$
and $I=\{1\leq i_{1},<\cdots,<i_{n}\leq N\}\subset\mathcal{L}$ the
correlation functions of $\eta_{t}$ have bi-Pfaffian form
\begin{equation}
\rho_{n}(i_{1},\dots,i_{n};t)=\frac{1+\Phi}{2}\mathrm{Pf}\left.\mathbf{K}^{+}_{t}\right|_{I\times I}+\frac{1-\Phi}{2}\mathrm{Pf}\left.\mathbf{K}^{-}_{t}\right|_{I\times I},\label{eq: rho_n bi-Pfaffian}
\end{equation}
if the same holds for the initial measure $\mu_{0}$. Here, $\mathbf{K}^{\pm}_{t}$
are matrix kernels, $\mathbf{K}^{\pm}_{t}:(\mathcal{L\times}\{1,2\})\times(\mathcal{L\times}\{1,2\})\to\mathbb{R},$
i.e., skew-symmetric $2N\times2N$ block matrices with $2\times2$
blocks of the form (\ref{eq: 2 x 2 blocks}). The blocks are obtained
``in the derived form'' from the corresponding ``scalar kernels'' $K^{\pm}_{t}:\{i,j:1\leq i\leq j\leq N\}\to\mathbb{R}$
as follows. 
\[
\mathbf{K}^{\pm}_{t}(i,j)=\frac{1}{1+\theta}\left(\begin{array}{cc}
K^{\pm}_{t}(i,j) & -\nabla^{+}_{2}K^{\pm}_{t}(i,j)\\
-\nabla^{+}_{1}K^{\pm}_{t}(i,j) & \nabla^{+}_{1}\nabla^{+}_{2}K^{\pm}_{t}(i,j)
\end{array}\right),\quad i<j,
\]
 and 
\[
\mathbf{K}^{\pm}_{t,12}(i,i)=-(1+\theta)^{-1}\nabla^{+}_{2}K^{\pm}_{t}(i,i),
\]
while the other matrix coefficients are determined by the skew symmetry.
Here, $\nabla^{+}_{\cdot}$ is the discrete derivative,
\[
\nabla^{+}f(x):=f(x+1)-f(x),
\]
with a subscript that specifies the argument $\nabla^{+}_{\cdot}$
acts on. The way the differentiation respects the cyclic structure
of $\mathbb{Z}_{N}$ will be clear from the periodic boundary conditions (BC)
for $K^{\pm}_{t}$ below. 

The scalar kernels $K^{\pm}_{t}(x,y)$ are defined for $1\leq x\leq y\leq N$
to be the unique solutions to the ``drifted diffusion equation'',
\begin{equation}
\frac{d}{dt}K^{\pm}_{t}(x,y)=\left(\Delta^{r,l}_{1}+\Delta^{r,l}_{2}\right)K^{\pm}_{t}(x,y),\quad x<y,\label{eq:  diffusion equation for K}
\end{equation}
supplied with BC
\begin{equation}
K^{\pm}_{t}(x,x)=1,\label{eq: BC K(xx)}
\end{equation}
periodic BC 
\begin{align}
K^{\pm}_{t}(x,y) & =\pm K^{\pm}_{t}(y,x+N),\label{eq: periodic BC K}
\end{align}
and initial conditions 
\begin{equation}
K^{\pm}_{0}(x,y)=\frac{\mathbb{E}\left[\left(-\theta\right)^{\eta_{0}[x,y)}\pm\left(-\theta\right)^{|\eta_{0}|-\eta_{0}[x,y)}\right]}{\left(1\pm\Phi\right)},\label{eq: initial cond K}
\end{equation}
 where $\eta[x,y)=\sum^{y-1}_{i=x}\eta(i)$ for any $\eta\in\Omega$,
and $\Delta^{r,l}$ is ``drifted Laplacian'' acting on functions on
$\mathcal{L}$ as 
\begin{equation}
\Delta^{r,l}f(x)=rf(x-1)+lf(x+1)-f(x).\label{eq: drifted Laplacian}
\end{equation}
 The subscripts of the operators specify the argument they act on.
The above BC are to be used when the action of the Laplacians or the
discrete derivatives takes the argument outside of the specified domain
$1\leq x\leq y\leq N$.
\end{thm}
\begin{rem}
\label{rem:Phi=00003Dpm 1}As $(1\pm\Phi)$ appears in the denominator,
the cases $\Phi=\pm1$ should be treated with care. Thus, we excluded these cases from the theorem statement and fill the gap in this remark.
For $\theta<1$, one of the identities, $\Phi=1$, holds in and only
in the trivial case $\mathbb{P}(|\eta_{0}|=0)=1$ of the initially
empty lattice, which implies that all the correlation functions vanish
identically at arbitrary time. The second identity, $\Phi=-1$, never
holds in this case. In the pure annihilation case $\theta=1$, the
identity $\Phi=1$ ($\Phi=-1$) holds if and only if the initial distribution
$\mu_{0}$ is concentrated on the subset of $\Omega$ consisting of
the particle configurations with fixed parity, i.e. $\mathbb{P}(|\eta_{0}|\in2\mathbb{Z})$=1
($\mathbb{P}(|\eta_{0}|\in2\mathbb{Z}+1)$=1). Provided the
initial measure satisfies the corresponding single-Pfaffian condition,
only the term in (\ref{eq: rho_n bi-Pfaffian}) with a non-zero
prefactor remains, yielding a single-Pfaffian form of the correlation
functions. Then, the measure on configurations is a pure PPP. 
\end{rem}
\begin{rem}
When $\theta=0$, i.e. in the case of pure coalescence, $(-\theta)^{n}$
should be understood as $\mathbb{I}_{n=0}$. In particular, 
\[
\Phi=\mathbb{P}(|\eta_{0}|=0)
\]
 and 
\[
K^{\pm}_{0}(x,y)=\frac{\mathbb{P}(\eta_{0}[x,y)=0)\pm\mathbb{P}\left(\eta_{0}[x,y)=|\eta_{0}|\right)}{1\pm\Phi}.
\]
\end{rem}
Theorem \ref{thm: bi-PPP} extends the result of \cite{garrod2018examples}
obtained for CARW on $\mathcal{L}=\mathbb{Z}$ to the case of the
ring $\mathcal{L}=\mathbb{Z}_{N}$. While in the former case the measure
on $\eta_{t}$ is a PPP, in the latter the correlation functions are
given by the linear combination of two Pfaffians. Hence, it is natural
to refer to the corresponding point process as bi-Pfaffian. To our knowledge, point processes of this type have not yet been considered.

Initial conditions with correlation functions having the bi-Pfaffian
structure required by the theorem \ref{thm: bi-PPP} may look restrictive
at first glance. However, below we prove that this structure takes
place at least for a class of initial conditions given by the inhomogeneous
Bernoulli point process, with occupation numbers of all sites being
independent random variables. A particular limiting case is the initial
distribution concentrated on an arbitrary single configuration.
\begin{lem}
\label{lem: Inhomogeneouse Bernoulli IC}Let $\boldsymbol{p}=(p_{1},\dots,p_{N})$
be an $N$-tuple of probabilities, $0\leq p_{i}\leq1,1\leq i\leq N,$
and $\mu{}_{Ber(\boldsymbol{p};N)}$ be the inhomogeneous Bernoulli
point process on $\Omega=\{0,1\}^{\mathcal{L}}$ with $\mathcal{L}=\{1,\dots,N\},$
i.e., the probability measure assigning probability 
\[
\mu{}_{Ber(\boldsymbol{p};N)}(\eta)=\prod^{N}_{i=1}p^{\eta(i)}_{i}\left(1-p_{i}\right)^{1-\eta(i)}
\]
to a configuration $\eta\in\Omega.$ Then, the correlation functions
of $\mu{}_{Ber(\boldsymbol{p};N)}(\eta)$ have the bi-Pfaffian form
(\ref{eq: rho_n bi-Pfaffian}). 
\end{lem}
\begin{rem}
In particular, the lemma holds in the limiting case of the fixed initial
configuration, i.e. with $\boldsymbol{p}=\eta'$ for any $\eta'\in\Omega$,
which implies 
\[
\mu_{Ber(\eta';N)}(\eta)=\delta_{\eta,\eta'}.
\]

Below, in Section \ref{sec:Solution}, we also construct the solution
to eqs. (\ref{eq:  diffusion equation for K}--\ref{eq: initial cond K})
as an expansion in explicitly defined eigenvectors of $\Delta^{r,l}$.
To avoid overloading the article with formulas, we do not construct
any explicit solutions for concrete examples of initial conditions
here. This is left for future work. In particular, it would be interesting
to compare the results with those on $\mathbb{Z}$. As was noted,
the decay of the density and other correlation functions with time
is algebraic in the infinite system, whereas on the ring the density
converges exponentially to $\frac{1-\Phi}{N(1+\theta)}$ and
correlations of order at least two decay exponentially.
It would be interesting to describe the crossover between these two
types of behaviour. A more ambitious problem is to evaluate also the
unequal-time correlation functions. Whether they lead to a bi-Pfaffian
analogue of the extended Pfaffian process at least in the case $\theta=1$
remains an open question. 
\end{rem}

\section{Markov duality and correlation functions\label{sec:Annihilating-random-walk}}

In this section, we give a proof of Theorem \ref{thm: bi-PPP} using
the Markov duality method in the form outlined in \cite{povolotsky2025representations}.
First, we note that the evolution of the expectation of a function $f:\Omega\to\mathbb{R}$
of configuration at time $t$ is governed by the backward Kolmogorov
equation 
\begin{equation}
\frac{d}{dt}\mathbb{E}f(\eta_{t})=\mathbb{E}Lf(\eta_{t}),\label{eq: Kolmogorov eq.}
\end{equation}
where the action of the generator $L$ is defined by
\begin{equation}
Lf(\eta)=\sum_{\eta'\in\Omega}u_{\eta\eta'}(f(\eta')-f(\eta)),\quad\eta\in\Omega\label{eq: Generator}
\end{equation}
with $u_{\eta,\eta'}$ for $\eta,\eta'\in\Omega$ being the rate of
transition from configuration $\eta$ to configuration $\eta'.$ All
these transitions change the state of a pair of neighboring sites,
so that the sum on the r.h.s. of (\ref{eq: Kolmogorov eq.}) can in fact
be constructed as an external sum over all such pairs and an internal
sum over all possible transitions from configuration $\eta$ due to a change of a given pair, where all the possibilities are exhausted
by the list (\ref{eq: rates}).

Note that the evolution of correlation functions themselves being
expectations of products of occupation numbers of given sites is governed
by equations of the form (\ref{eq: Kolmogorov eq.}), which, however,
constitute a hardly tractable system of coupled difference-differential
equations as large as the total number of all correlation functions.
The idea of the subsequent Markov-duality-based procedure is to construct
an alternative set of functions such that the correlation functions
could be reconstructed from them, and the system of equations could
be subdivided into smaller independent solvable subsystems. 

\subsection{Vector-function correspondence}

Instead of working directly with functions of configurations, it is
convenient to use an equivalent algebraic language. The space of functions $f:\Omega\to\mathbb{R}$ spanned by the basis set $\{\mathbb{I}_{\eta},\eta\in\Omega\}$
consisting of indicator functions, $\mathbb{I}_{\eta}:\eta'\mapsto\delta_{\eta,\eta'}$,
can be identified with a vector space $\mathcal{F}=\left(\mathbb{R}^{2}\right)^{\otimes\mathcal{L}}$
by identifying the basis vectors

\[
\mathbb{I}_{\eta}\mapsto\boldsymbol{e}^{(\eta)}:=\bigotimes^{N}_{i=1}e^{(\eta_{i})}_{i},\quad\eta\in\Omega.
\]
Here, $e^{(0)}_{i},e^{(1)}_{i}$ are copies of the basis vectors of
$\mathbb{R}^{2}$ associated with the factor $i\in\mathcal{L}$ of
the tensor product,
\[
e^{(0)}=\left(\begin{array}{c}
0\\
1
\end{array}\right),\quad e^{(1)}=\left(\begin{array}{c}
1\\
0
\end{array}\right).
\]
They correspond to indicators of zero and one at site $i$, respectively.
This correspondence establishes an isomorphism of linear spaces. In the following, we use the same notations, say $f(\cdot)$ and $f$,
for functions and for their vector counterparts, adding the argument
in parentheses to the former. 

Given a function $f(\eta)$ and the corresponding vector $f\in\mathcal{F}$, the expectation of $f(\eta_{t})$ can be represented as a scalar product
\begin{equation}
\mathbb{E}f(\eta_{t})=\langle\mu_{t},f\rangle\label{eq: expectation - scalar product}
\end{equation}
with a vector 
\[
\mu_{t}=\sum_{\eta\in\Omega}\mathbb{P}(\eta_{t}=\eta)\boldsymbol{e}^{(\eta)},
\]
where the basis $\{\boldsymbol{e}^{(\eta)}\}_{\eta\in\Omega}$ is
assumed to be orthonormal with respect to the scalar product $\langle\boldsymbol{e}^{(\eta)},\boldsymbol{e}^{(\eta')}\rangle=\delta_{\eta,\eta'}$.
Then, the Kolmogorov equation takes the form
\begin{equation}
\frac{d}{dt}\langle\mu_{t},f\rangle=\langle\mu_{t},Lf\rangle.\label{eq: Kolmogorov - vector form}
\end{equation}
The next step is to construct a set of vectors with particularly simple
action of $L$, which could be brought out of the scalar product,
so that the system would be reduced to a system of ODEs. 

\subsection{Duality functions}

As an operator acting in the tensor product space $\mathcal{F}$, the
generator is given by a sum
\[
L=\sum_{i\in\mathcal{L}}\ell_{i}
\]
 of two-site operators $\ell_{i}\in End(\mathcal{F})$, which act
as the matrix 
\[
\ell=\left(\begin{array}{cccc}
-1 & (1-\theta)l & (1-\theta)r & \theta\\
0 & -r & r & 0\\
0 & l & -l & 0\\
0 & 0 & 0 & 0
\end{array}\right)
\]
on the product of tensor factors $(i,i+1)$ with basis $(e^{(1)}\otimes e^{(1)},e^{(1)}\otimes e^{(0)},e^{(0)}\otimes e^{(1)},e^{(0)}\otimes e^{(0)})$
and identically on the other ones and $\ell_{N}$ acts non-trivially
on spaces $(N,1).$

The key observation of \cite{povolotsky2025representations} used
for the construction of Markov duality functions is as follows. Let us
introduce two column vectors 
\[
v=\left(\begin{array}{c}
1\\
1
\end{array}\right),w=\left(\begin{array}{c}
-\theta\\
1
\end{array}\right).
\]
The matrix $\ell$ acting in the tensor square $\mathbb{R}^{2}\otimes\mathbb{R}^{2}$
of the two-dimensional spaces ``kills'' the tensor squares of $v$
and $w$,

\begin{equation}
\ell v\otimes v=0,\quad\ell w\otimes w=0,\label{eq: ell vv ell ww}
\end{equation}
and acts on the mixed products as follows
\begin{align}
\ell v\otimes w & =rw\otimes w+lv\otimes v-v\otimes w,\quad\ell w\otimes v=rv\otimes v+lw\otimes w-w\otimes v.\label{eq: ell vw ell wv}
\end{align}
These formulas show that the action of the operators $\ell_{i}$,
and hence of the full generator $L$, on vectors in $\mathcal{F}$
given by tensor products of copies of $v$ and $w$, such as
\[
v_{1}\otimes\cdots\otimes v_{x_{1}-1}\otimes w_{x_{1}}\otimes\cdots\otimes w_{x_{2}-1}\otimes v_{x_{2}}\otimes\cdots\otimes w_{x_{2n}-1}\otimes v_{x_{2n}}\cdots\otimes v_{N},
\]
is very simple. Its only effect is on the ``domain walls'' separating
the vectors $v$ and $w$ in two adjacent tensor product factors.
The action of $L$ results in a linear combination of vectors of the
same type, where one of the domain walls shifts right or left, or stays put with coefficients $l$, $r,$ and $-1,$ respectively.
When two domain walls in adjacent positions, as in $v\otimes w\otimes v$,
merge as a result of the shift of one of them, they simply disappear.
If one thinks of the initial tensor product as a sequence of alternating
clusters of vectors $v$ and $w$, the action of $L$ can only produce
the vectors obtained by shifting the cluster boundaries (domain walls),
under which the number of clusters can only decrease, when two boundaries
merge, and no new clusters appear. Hence, the resulting vectors consist
of the same or a smaller number of clusters. 

Obviously, up to interchange $v\leftrightarrow w$ the discussed tensor
product vectors can be identified solely in terms of positions of
the domain walls, and the r.h.s. of the evolution equations for the
corresponding functions contains only the functions corresponding to
other vectors with the same number of domain walls or fewer. Thus, these equations
can be solved recursively by increasing the number of domain walls.
This was the main idea used in \cite{garrod2018examples} to solve
CARW on $\mathbb{Z}$. Here we notice that the Kolmogorov equations
written for the expectations of the functions thus constructed, indexed
by the domain wall positions, look exactly like the Kolmogorov equations
for different functions of configurations of a ``dual Markov process'',
which is the space-reversed ARW with a finite number of particles.
There, the domain walls play the role of the particles in the dual
process and ``space reversed'' means that the rates $l$ and $r$
are swapped with respect to the original process. Hence, the concept
of ``Markov duality'' turns out to be relevant. We do not return to
this point in our article, though exploit the ``duality function''
term below. See \cite{povolotsky2025representations} for further
details.

Note that in the case of the infinite integer lattice, the tensor product
space was the span of vectors with at most a finite number of $w$-factors,
and corresponded to the space of functions depending on the state
of a finite number of sites only. Our case of a finite periodic system,
where the vectors $v$ and $w$ are in a sense on an equal footing, requires
a modification of this scheme, which would respect the periodic BC. 

To this end, we introduce a pair of ``complementary'' duality functions
as follows. Let us fix $\boldsymbol{x}:=(x_{1},\dots,x_{2n})\in D_{2n}$
where
\begin{equation}
D_{n}=\{\boldsymbol{x}:1\leq x_{1}<\dots<x_{n}\leq N\},\quad n\geq1.\label{eq: D_n}
\end{equation}
 and define a pair of vectors from $\mathcal{F}$
\begin{align}
f^{(2n)}_{\boldsymbol{x};\theta} & :=v_{1}\otimes\cdots\otimes v_{x_{1}-1}\otimes w_{x_{1}}\otimes\cdots\otimes w_{x_{2}-1}\otimes v_{x_{2}}\otimes\cdots\otimes w_{x_{2n}-1}\otimes v_{x_{2n}}\cdots\otimes v_{N},\label{eq: f}\\
\bar{f}^{(2n)}_{\boldsymbol{x};\theta} & :=w_{1}\otimes\cdots\otimes w_{x_{1}-1}\otimes v_{x_{1}}\otimes\cdots\otimes v_{x_{2}-1}\otimes w_{x_{2}}\otimes\cdots\otimes v_{x_{2n}-1}\otimes w_{x_{2n}}\cdots\otimes w_{N}.\label{eq: f-bar}
\end{align}
The corresponding functions count particles in $n$ clusters $[x_{1},x_{2}-1]\bigcup\cdots\bigcup[x_{2n-1},x_{2n}-1]$
and $n$ complementary clusters $\mathcal{L}\backslash[x_{1},x_{2}-1]\bigcup\cdots\bigcup[x_{2n-1},x_{2n}-1]$
respectively, assigning weight $(-\theta)$ to every particle, 
\begin{align}
f^{(2n)}_{\boldsymbol{x};\theta} & (\eta)=\left(-\theta\right)^{\sum^{n}_{i=1}\eta[x_{2i-1},x_{2i})}\label{eq: f-fun}\\
\bar{f}^{(2n)}_{\boldsymbol{x};\theta} & (\eta)=\left(-\theta\right)^{|\eta|-\sum^{n}_{i=1}\eta[x_{2i-1},x_{2i})}.\label{eq: f-fun-bar}
\end{align}
Correspondingly, for their expectations at time $t$ we introduce
functions of positions of domain walls
\begin{align}
\mathfrak{f}{}^{(2n)}_{t}(\boldsymbol{x}) & :=\mathbb{E}f^{(2n)}_{\boldsymbol{x};\theta}(\eta_{t}),\label{eq: f_t^(2n)(x;theta)}\\
\bar{\mathfrak{f}}{}^{(2n)}_{t}(\boldsymbol{x}) & :=\mathbb{E}\bar{f}^{(2n)}_{\boldsymbol{x};\theta}(\eta_{t}).\label{eq: bar f_t^(2n)(x;theta)}
\end{align}
These functions can be found as a solution of the system of evolution
equations given in the following lemma.
\begin{lem}
For $\bm{x}\in D_{2n}$ functions (\ref{eq: f_t^(2n)(x;theta)} and \ref{eq: bar f_t^(2n)(x;theta)})
satisfy equations 
\begin{equation}
\partial_{t}\mathfrak{f}^{(2n)}_{t}(\boldsymbol{x})=\Delta^{(2n)}\mathfrak{f}^{\left(2n\right)}_{t}(\boldsymbol{x}),\quad\partial_{t}\mathfrak{\bar{f}}^{(2n)}_{t}(\boldsymbol{x})=\Delta^{(2n)}\mathfrak{\bar{f}}^{\left(2n\right)}_{t}(\boldsymbol{x}),\quad n\geq1\label{eq: evolution equations}
\end{equation}
where 
\[
\Delta^{(2n)}=\sum^{2n}_{i=1}\Delta^{r,l}_{i}
\]
with $\Delta^{r,l}_{i}$ being the drifted Laplace operator (\ref{eq: drifted Laplacian})
acting on the variable $x_{i}$, $i=1,\dots,2n.$ These equations are
supplied with BC used to replace the terms with arguments outside of
the domain $D_{2n}$, appearing as a result of the action of $\Delta^{(2n)}$
on the r.h.s., by the allowed ones. The BC
\begin{align}
\mathfrak{f}^{(2n)}_{t}(\dots,x,x\dots) & =\mathfrak{f}^{(2n-2)}_{t}(\dots,\widehat{x,x}\dots),\quad n>1,\label{eq: f(...xx...) BC}\\
\mathfrak{\bar{f}}^{(2n)}_{t}(\dots,x,x\dots) & =\mathfrak{\bar{f}}^{(2n-2)}_{t}(\dots,\widehat{x,x}\dots),\quad n>1,\label{eq: bar  f(...xx...) BC}
\end{align}
 where the hat $\widehat{\,\,\,}$ over arguments means that they
are omitted, and
\begin{align}
\mathfrak{f}^{(2)}_{t}(x,x) & =1,\label{eq: f BC joining n=00003D2}\\
\mathfrak{\bar{f}}^{(2)}_{t}(x,x) & =\Phi=\mathbb{E}\left(-\theta\right)^{|\eta_{0}|}.\label{eq:f bar BC joining n=00003D2}
\end{align}
are used in cases when two subsequent arguments coincide, and those
\begin{align}
\mathfrak{f}{}^{(2n)}_{t}(x_{1},\dots,x_{2n}) & =\bar{\mathfrak{f}}{}^{(2n)}_{t}(x_{2},\dots,x_{2n},x_{1}+N),\label{eq: periodic BC f vs bar f}\\
\bar{\mathfrak{f}}{}^{(2n)}_{t}(x_{1},\dots,x_{2n}) & =\mathfrak{f}{}^{(2n)}_{t}(x_{2},\dots,x_{2n},x_{1}+N).\label{eq: Periodic BC bar f vs f}
\end{align}
take the ring periodicity into account.
\end{lem}
\begin{proof}
Equations (\ref{eq: evolution equations}) directly follow from the
scalar product representation of expectations (\ref{eq: expectation - scalar product})
and the Kolmogorov equation (\ref{eq: Kolmogorov - vector form}) and
from the form (\ref{eq: ell vv ell ww} and \ref{eq: ell vw ell wv})
of the action of the operators $\ell_{i}$ on the tensor squares and
cross-products of vectors $v$ and $w$. As was already noted, the
action of $L$ on vectors of the form (\ref{eq: f} and \ref{eq: f-bar})
results in linear combinations of vectors of the same form with shifted
$\bm{x}$-coordinates and, hence, can be taken out of the scalar product,
resulting in (\ref{eq: evolution equations}). 

BC (\ref{eq: f(...xx...) BC} and \ref{eq: bar  f(...xx...) BC}) simply
state that when two domain walls merge and disappear after a move
of one of them, the cluster that was between them disappears, while
the other cluster boundaries remain unchanged. Analogous conditions
(\ref{eq: f BC joining n=00003D2} and \ref{eq:f bar BC joining n=00003D2})
for the case $n=1$ treat two possible situations in which no domain
walls remain in the tensor product. In this case, the vectors $f^{(2)}_{\boldsymbol{x};\theta}$
and $\bar{f}^{(2)}_{\boldsymbol{x};\theta}$ with $\bm{x}=(x,x),1\leq x\leq N$,
should be identified with 
\[
f^{(2)}_{\boldsymbol{x};\theta}:=\otimes_{k\in\mathcal{L}}v_{k},\quad\bar{f}^{(2)}_{\boldsymbol{x};\theta}:=\otimes_{k\in\mathcal{L}}w_{k}.
\]
The first one corresponds to the function $f^{(2)}_{\boldsymbol{x};\theta}(\eta)\equiv1,$
while the second is associated with the function
\[
\bar{f}^{(2)}_{\boldsymbol{x};\theta}(\eta)=\phi(\eta):=\left(-\theta\right)^{|\eta|}.
\]

The expectations of both functions are time independent. Indeed, the
evolution of functions (\ref{eq: f-fun} and \ref{eq: f-fun-bar})
is associated with the motion of domain walls in the corresponding
tensor product. However, there are no domain walls in the corresponding
vectors. Of course, the time independence is trivially obvious for
the constant function. For $\phi$ we also have 
\[
\mathbb{E}\phi(\eta_{t})=\mathbb{E}\phi(\eta_{0}),\quad t>0.
\]
In other words, $\phi(\eta_{t})$ is a martingale. Thus, the l.h.s.
of the identity (\ref{eq:f bar BC joining n=00003D2}) is in fact
time independent. 

Finally, the periodic BC (\ref{eq: periodic BC f vs bar f} and \ref{eq: Periodic BC bar f vs f})
state that as soon as the first (last) coordinate of $\boldsymbol{x}$
becomes less than $1$ (greater than $N$), $\boldsymbol{x}$ should
be brought back to $D_{n}$ by adding to (subtracting from) this coordinate
the lattice period $N$. Simultaneously, the clusters of $v$-s and
$w$-s should be swapped to maintain consistency with the definitions
of vectors (\ref{eq: f} and \ref{eq: f-bar}).
\end{proof}

\begin{rem}
Identities (\ref{eq: f(...xx...) BC}--\ref{eq:f bar BC joining n=00003D2})
allow one to extend the domain of functions $\mathfrak{f}{}^{(2n)}_{t}(\bm{x})$
and $\bar{\mathfrak{f}}{}^{(2n)}_{t}(\bm{x})$ from $D_{2n}$ defined
in (\ref{eq: D_n}) to $\tilde{D}_{2n}$, where
\[
\tilde{D}_{n}=\{\boldsymbol{x}:x_{1}\leq\dots\leq x_{n}\leq x_{1}+N\},\quad n\geq1.
\]
This extension is also used when $2n>N$, so that $D_{2n}$ is empty.

It is convenient to reformulate the problem as follows. Let us introduce
notations $\mathfrak{f}{}^{(2n)}_{t,+}$ and $\mathfrak{f}{}^{(2n)}_{t,-}$
for symmetric and skew-symmetric combinations of $f^{(2n)}_{\boldsymbol{x};\theta}$
and $\bar{f}^{(2n)}_{\boldsymbol{x};\theta}$ 
\begin{align*}
\mathfrak{f}{}^{(2n)}_{t,\pm}(\boldsymbol{x}) & =\mathfrak{f}{}^{(2n)}_{t}(\boldsymbol{x})\pm\bar{\mathfrak{f}}{}^{(2n)}_{t}(\boldsymbol{x}),
\end{align*}
Obviously, the new functions solve the same equation 
\[
\partial_{t}\mathfrak{f}{}^{(2n)}_{t,\pm}(\boldsymbol{x})=\Delta^{(2n)}\mathfrak{f}{}^{(2n)}_{t,\pm}(\boldsymbol{x}),
\]
as $\mathfrak{f_{t}}^{(2n)}$ and $\mathfrak{\bar{f}}^{(2n)}_{t}$
do, while the BC become 
\begin{align}
\mathfrak{f}^{(2n)}_{t,\pm}(\dots,x,x\dots) & =\mathfrak{f}^{(2n-2)}_{t,\pm}(\dots,\widehat{x,x}\dots),n>1\label{eq: BC fpm(xx) n>1}\\
\mathfrak{f}^{(2)}_{t,\pm}(x,x) & =1\pm\Phi\label{eq: BC fpm(xx) n=00003D2}\\
\mathfrak{f}{}^{(2n)}_{t,\pm}(x_{1},\dots,x_{2n}) & =\pm\mathfrak{f}{}^{(2n)}_{t,\pm}(x_{2},\dots,x_{2n},x_{1}+N),\label{eq: BC fpm periodic}
\end{align}
 i.e. we reduce the problem to problems with periodic and anti-periodic BC. The solution can be sought in Pfaffian form.
\end{rem}
\begin{lem}
\label{lem:Pfaffian theta}Suppose that $\Phi\neq\pm1$ and at $t=0$
relations 
\begin{equation}
\mathfrak{f}{}^{(2n)}_{t,\pm}(x_{1},\dots,x_{2n})=(1\pm\Phi)\mathrm{Pf}\left(\frac{\mathfrak{f}^{(2)}_{t,\pm}(x_{i},x_{j})}{1\pm\Phi};1\leq i<j\leq2n\right)\label{eq:Pfaffian theta}
\end{equation}
hold. Then they hold at any $t>0$. 
\end{lem}
\begin{proof}
The check that the r.h.s. satisfies (\ref{eq: evolution equations})
and (\ref{eq: f(...xx...) BC} and \ref{eq: bar  f(...xx...) BC})
is the same as in the case of the infinite lattice and directly follows
from the Pfaffian definition (\ref{eq: Pfaffian def}) (see \cite{garrod2018examples}).
The cyclic BC (\ref{eq: BC fpm periodic}) also hold: under cyclic
reordering, the Pfaffian picks up the sign of the $(2n)$-cycle, while
the two-point periodicity supplies the compensating boundary sign,
giving precisely the factor $\pm$ in (\ref{eq: BC fpm periodic}).
Uniqueness follows by induction on the number of domain walls: for fixed
$N$, each level is a finite-dimensional linear system, with boundary
terms already determined by the preceding levels.
\end{proof}

Now, we are in a position to finish the proof of Theorem \ref{thm: bi-PPP}.
\begin{proof}
(of Theorem \ref{thm: bi-PPP})

Let $\bm{x}=\{x_{1},\dots,x_{n}\}\in D_{n}.$ To obtain the result,
we set 
\[
K^{\pm}_{t}(x,y)=\frac{\mathfrak{f}^{(2)}_{t,\pm}(x,y)}{1\pm\Phi},\quad1\leq x\leq y\leq N.
\]
Then, the formula for the correlation functions follows from the identity
\[
s=\frac{1-\left(-\theta\right)^{s}}{1+\theta},\quad s\in\{0,1\},
\]
which implies 
\[
\rho_{n}(x_{1},\dots,x_{n};t)=\left(-\frac{\nabla^{+}_{y_{1}}}{1+\theta}\right)_{y_{1}=x_{1}}\cdots\left(-\frac{\nabla^{+}_{y_{n}}}{1+\theta}\right)_{y_{n}=x_{n}}\mathfrak{f}^{(2n)}_{t}(x_{1},y_{1},\dots,x_{n},y_{n}),
\]
where we use BC (\ref{eq: f(...xx...) BC}). Reconstructing $\mathfrak{f}^{(2n)}_{t}(x_{1},y_{1},\dots,x_{n},y_{n})$
as a half-sum of $\mathfrak{f}^{(2n)}_{t,+}$ and $\mathfrak{f}^{(2n)}_{t,-}$
written in the Pfaffian form (\ref{eq:Pfaffian theta}) and applying
the discrete derivatives to them yields the linear combination (\ref{eq: rho_n bi-Pfaffian})
of Pfaffians of the block matrices. Obviously, the Pfaffian formulas
(\ref{eq:Pfaffian theta}) of initial conditions for $\mathfrak{f}^{(2n)}_{t,+}$
and $\mathfrak{f}^{(2n)}_{t,-}$ and (\ref{eq: rho_n bi-Pfaffian})
for the correlation functions are equivalent.
\end{proof}

\begin{rem}
In the same way one obtains the correlation functions of the form
$\mathbb{E}\eta_{t}(x_{1})\cdots\eta_{t}(x_{n})(-\theta)^{|\eta_{t}|}$
from the function given by the half-difference of $\mathfrak{f}^{(2n)}_{t,+}$
and $\mathfrak{f}^{(2n)}_{t,-}$.
\end{rem}
\begin{rem}
According to Remark \ref{rem:Phi=00003Dpm 1}, the nontrivial cases
$\Phi=\pm1$ are possible exactly in the case $\theta=1$ of pure
ARW, when the initial number of particles is restricted to be either
odd or even. Then, one of the two functions $\mathfrak{f}{}^{(2n)}_{t,\pm}$
vanishes, and formula (\ref{eq:Pfaffian theta}) is applied to the
other.
\end{rem}

\section{Initial conditions}

Now, the question is which initial conditions satisfy the conditions of Lemma \ref{lem:Pfaffian theta}. The following lemma shows that
at least the general inhomogeneous Bernoulli point process is in this
class. Since the duality functions and the scalar kernels $K^{\pm}_{t}$
are in one-to-one correspondence, the statement of Lemma \ref{lem: Inhomogeneouse Bernoulli IC}
is its immediate corollary. 
\begin{lem}
Let $\mu_{Ber(\boldsymbol{p})}(\eta)$ be an inhomogeneous Bernoulli
point process on $\mathcal{L}=\{1,\dots,N\}$ with parameters $\boldsymbol{p}=\{p_{i}:0\leq p_{i}\leq1,1\leq i\leq N\}$
defined in the statement of Lemma \ref{lem: Inhomogeneouse Bernoulli IC}.
For $\bm{x}\in D_{2n}$ and $n\geq1$ define 
\[
\mathfrak{f}^{(2n)}(\bm{x}):=\mathbb{E}\left(-\theta\right)^{\sum^{n}_{i=1}\eta[x_{2i-1},x_{2i})},\quad\bar{\mathfrak{f}}^{(2n)}(\bm{x}):=\mathbb{E}\left(-\theta\right)^{|\eta|-\sum^{n}_{i=1}\eta[x_{2i-1},x_{2i})}
\]
and 
\[
\Phi:=\mathbb{E}\left(-\theta\right)^{|\eta|}.
\]
We assume $\Phi\neq\pm1$.

Then, 
\begin{eqnarray}
\frac{\mathfrak{f}^{(2n)}(\bm{x})\pm\bar{\mathfrak{f}}^{(2n)}(\bm{x})}{1\pm\Phi} & = & \mathrm{Pf}\left(\frac{\mathfrak{f}^{(2)}(x_{i},x_{j})\pm\bar{\mathfrak{f}}^{(2)}(x_{i},x_{j})}{1\pm\Phi},1\leq i<j\leq2n\right),\label{eq: claim}
\end{eqnarray}
for $1\leq n\leq N$, interpreting the identity on $\tilde{D}_{2n}$
when $2n>N$ via the collision and cyclic boundary relations. 

In the case $\theta=0$, the function $\left(-\theta\right)^{n},$$n\in\mathbb{N}_{0}$
should be understood as $\mathbb{I}_{n=0}$.
\end{lem}
\begin{proof}
Assume first that none of the identities 
\begin{equation}
1-p_{i}(1+\theta)=0,\quad i=1,\dots,N,\label{eq: 1-(p_=00007Bi=00007D+1)theta=00003D0}
\end{equation}
holds. This guarantees that $\mathfrak{f}^{(2n)}(\bm{x})\neq0,\bar{\mathfrak{f}}^{(2n)}(\bm{x})\neq0,$
for any $\boldsymbol{x}\in D_{2n}$ and any $n\geq1$. Due to the independence of the states of different sites in the Bernoulli point process, we have
\begin{eqnarray}
\mathfrak{f}^{(2n)}(x_{1},\ldots,x_{2n}) & = & \prod^{n}_{k=1}\mathfrak{f}^{(2)}(x_{2k-1},x_{2k}).\label{2n2}
\end{eqnarray}
Also, since $\mathfrak{f}^{(2n)}(\bm{x})$ and $\bar{\mathfrak{f}}^{(2n)}(\bm{x})$
are expectations of functions depending on the independent complementary
sets of sites, their product yields $\Phi$,
\begin{eqnarray}
\bar{\mathfrak{f}}^{(2n)}(x_{1},\ldots,x_{2n})=\frac{\Phi}{\mathfrak{f}^{(2n)}(x_{1},\ldots,x_{2n})} & \stackrel{(\ref{2n2})}{=} & \frac{\Phi}{\prod^{n}_{k=1}\mathfrak{f}^{(2)}(x_{2k-1},x_{2k})}.\label{eq: fbf}
\end{eqnarray}
Using (\ref{2n2},\ref{eq: fbf}) in (\ref{eq: claim}), we conclude
that the statement we need to prove is 
\begin{eqnarray}
 &  & \frac{\prod^{n}_{k=1}\mathfrak{f}^{(2)}(x_{2k-1},x_{2k})\pm\Phi\left(\prod^{n}_{k=1}\mathfrak{f}^{(2)}(x_{2k-1},x_{2k})\right)^{-1}}{1\pm\Phi}\nonumber \\
 &  & \,\,\,\,\,=\mathrm{Pf}\left(\frac{\mathfrak{f}^{(2)}(x_{i},x_{j})\pm\Phi(\mathfrak{f}^{(2)}(x_{i},x_{j}))^{-1}}{1\pm\Phi},1\leq i<j\leq2n\right),n\in\mathbb{N}.\label{claim1}
\end{eqnarray}
Define $a(x)=\mathfrak{f}^{(2)}(1,x)=\mathbb{E}(-\theta)^{\eta[1,x)}$,
$x=1,\dots,N$. Then, $\mathfrak{f}^{(2)}(x,y)=\frac{a(y)}{a(x)}$,
$1\leq x<y\leq N$ and one observes that the claim (\ref{claim1})
follows from a more general claim 
\begin{eqnarray}
 &  & \frac{1}{1+\Phi}\left(\frac{a_{2}}{a_{1}}\frac{a_{4}}{a_{3}}\ldots\frac{a_{2n}}{a_{2n-1}}+\Phi\frac{a_{1}}{a_{2}}\frac{a_{3}}{a_{4}}\ldots\frac{a_{2n-1}}{a_{2n}}\right)\label{claim2}\\
 &  & \,\,\,\,\,\,\,\,\,=\mathrm{Pf}\left(\frac{1}{1+\Phi}\left(\frac{a_{j}}{a_{i}}+\Phi\frac{a_{i}}{a_{j}}\right),1\leq i<j\leq2n\right),\nonumber 
\end{eqnarray}
for any $\Phi\in\mathbb{C}\setminus\{-1\}$, $a_{1},a_{2},\ldots,a_{2n}\in\mathbb{C}\setminus{0}$.
Multiplying both sides of (\ref{claim2}) by $\prod^{2n}_{k=1}a_{k}^{-1}$,
setting $\alpha_{k}=a_{k}^{-2}$, and using the identity $\mathrm{Pf}(BAB^{T})=\det(B)\mathrm{Pf}(A)$,
one concludes that (\ref{claim2}) is implied by 

\begin{eqnarray}
\frac{\left(\alpha_{1}\alpha_{3}\ldots\alpha_{2n-1}+\Phi\alpha_{2}\alpha_{4}\ldots\alpha_{2n}\right)}{1+\Phi}=\mathrm{Pf}\left(\frac{\alpha_{i}+\Phi\alpha_{j}}{1+\Phi},1\leq i<j\leq2n\right),\label{claim3}
\end{eqnarray}
where $\Phi\in\mathbb{C}\setminus\{-1\},\alpha_{k}\in\mathbb{C},1\leq k\leq2n$,
$n\in\mathbb{N}$. This is the statement which will be proved. Below
we give a proof based on the use of Grassmann variables and the Berezin integral representation of the Pfaffian. A simpler inductive proof will
be given in the forthcoming article \cite{povolotsky2026coalescing}.\\

To prove (\ref{claim3}), define $P_{n}$ to be its right-hand side.
Using the representation of the Pfaffian as a Berezin integral, one finds:
\begin{eqnarray}
P_{n}=\int_{\mathbb{R}^{0|2n}}d\theta_{2n\ldots1}\exp\left\{ \sum_{1\leq i<j\leq2n}\frac{\alpha_{i}+\Phi\alpha_{j}}{1+\Phi}\theta_{i}\theta_{j}\right\} ,
\end{eqnarray}
where $d\theta_{2n\ldots1}:=d\theta_{2n}d\theta_{2n-1}\ldots d\theta_{1}$
and the integral is understood as an iterated integral. Let us define
a bosonic variable and a pair of fermionic variables: 
\begin{eqnarray*}
E_{m}: & = & \exp\left\{ \sum_{m<i<j\leq2n}\frac{\alpha_{i}+\Phi\alpha_{j}}{1+\Phi}\theta_{i}\theta_{j}\right\} ,\\
\phi_{m}: & = & \sum_{m<j\leq2n}\theta_{j},\\
\psi_{m}: & = & \sum_{m<j\leq2n}\alpha_{j}\theta_{j},
\end{eqnarray*}
where $0\leq m\leq2n$. Integrating over $\theta_{2},\theta_{1}$, we obtain
\begin{eqnarray}
P_{n} & = & \int_{\mathbb{R}^{0|2n}}d\theta_{2n\ldots1}e^{\theta_{1}\theta_{2}\frac{\alpha_{1}+\Phi\alpha_{2}}{1+\Phi}}e^{\frac{\theta_{1}}{1+\Phi}(\alpha_{1}\phi_{2}+\Phi\psi_{2})}e^{\frac{\theta_{2}}{1+\Phi}(\alpha_{2}\phi_{2}+\Phi\psi_{2})}E_{2}\label{prec}\\
 & = & \frac{\alpha_{1}+\Phi\alpha_{2}}{1+\Phi}P_{n-1}+\frac{\Phi(\alpha_{2}-\alpha_{1})}{(1+\Phi)^{2}}J_{n-1},
\end{eqnarray}
where 
\begin{eqnarray}
J_{n-m}:=\int_{\mathbb{R}^{0\mid2n}}d\theta_{2n\ldots(2m+1)}\phi_{2m}\psi_{2m}E_{2m},0\leq m\leq n.
\end{eqnarray}
Integrating over the variables $\theta_{1},\theta_{2}$ in the integral
expression for $J_{n}$, one finds 
\begin{eqnarray}
J_{n}=(\alpha_{2}-\alpha_{1})P_{n-1}+\frac{\alpha_{2}+\Phi\alpha_{1}}{1+\Phi}J_{n-1}.\label{jrec}
\end{eqnarray}
Combining (\ref{prec}) and (\ref{jrec}), one arrives at the following recursion
for the pair $(P_{n},J_{n})$: 
\begin{eqnarray}
\left(\begin{array}{c}
P_{n}\\
J_{n}
\end{array}\right)=M_{1}\left(\begin{array}{c}
P_{n-1}\\
J_{n-1}
\end{array}\right)=M_{1}M_{2}\left(\begin{array}{c}
P_{n-2}\\
J_{n-2}
\end{array}\right)=\ldots=M_{1}M_{2}\ldots M_{n}\left(\begin{array}{c}
P_{0}\\
J_{0}
\end{array}\right),\label{rec}
\end{eqnarray}
where 
\begin{eqnarray}
\left(\begin{array}{c}
P_{0}\\
J_{0}
\end{array}\right)=\left(\begin{array}{c}
1\\
0
\end{array}\right),
\end{eqnarray}
and 
\begin{eqnarray*}
M_{k}=\left(\begin{array}{cc}
\frac{\alpha_{2k-1}+\Phi\alpha_{2k}}{1+\Phi} & \frac{(\alpha_{2k}-\alpha_{2k-1})\Phi}{(1+\Phi)^{2}}\\
\alpha_{2k}-\alpha_{2k-1} & \frac{\alpha_{2k}+\Phi\alpha_{2k-1}}{1+\Phi}
\end{array}\right),1\leq k\leq n.
\end{eqnarray*}
The matrix product $M_{1}M_{2}\ldots M_{n}$ is easy to calculate
explicitly, as the constituent matrices commute: define 
\begin{eqnarray*}
v_{1}:=\left(\begin{array}{c}
-1\\
1+\Phi
\end{array}\right),v_{2}:=\left(\begin{array}{c}
\Phi\\
1+\Phi
\end{array}\right).
\end{eqnarray*}
Then 
\begin{eqnarray}
M_{k}v_{1}=\alpha_{2k-1}v_{1};M_{k}v_{2}=\alpha_{2k}v_{2},1\leq k\leq n.\label{evs}
\end{eqnarray}
Notice that 
\begin{eqnarray}
\left(\begin{array}{c}
1\\
0
\end{array}\right)=\frac{v_{2}-v_{1}}{1+\Phi}.\label{ics}
\end{eqnarray}
Substituting (\ref{ics}) into (\ref{rec}) and calculating the product
of matrices using (\ref{evs}), one finds 
\begin{eqnarray*}
\left(\begin{array}{c}
P_{n}\\
J_{n}
\end{array}\right)=\left(\begin{array}{c}
\frac{\alpha_{1}\alpha_{3}\ldots\alpha_{2n-1}+\Phi\alpha_{2}\alpha_{4}\ldots\alpha_{2n}}{1+\Phi}\\
-\alpha_{1}\alpha_{3}\ldots\alpha_{2n-1}+\alpha_{2}\alpha_{4}\ldots\alpha_{2n}
\end{array}\right),n\in\mathbb{N},
\end{eqnarray*}
the first line of which proves (\ref{claim3}). 

Now we can extend the proof to the cases when $1-p_{i}(1+\theta)=0$
for some indices $i\in\{1,\dots,N\}$. Indeed, both the l.h.s.
and r.h.s. of (\ref{eq: claim}) are rational functions of the variable
$\theta$, coinciding for every $\theta\in[0,1]$ except for at most
$N$ points. Then, they coincide everywhere.
\end{proof}

\section{Solution for the scalar kernels \label{sec:Solution}}

We have reduced the problem of finding the equal-time correlation functions in the CARW model to the solution of eq. (\ref{eq:  diffusion equation for K})
subject to (\ref{eq: BC K(xx)}) and either periodic or antiperiodic
BC (\ref{eq: periodic BC K}) for the scalar kernels $K^{\pm}_{t}(x,y)$.
Let us look for them in the form 
\begin{equation}
K^{\pm}_{t}(x,y)=g^{\pm}_{t}(x,y)+r^{\pm}(x,y),\label{eq:g}
\end{equation}
where we introduce functions $g^{\pm}_{t}(x_{1},x_{2})$ solving (\ref{eq:  diffusion equation for K})
subject to (\ref{eq: periodic BC K}), while condition (\ref{eq: BC K(xx)})
is replaced by 
\begin{equation}
g^{\pm}_{t}(x,x)=0,\label{eq:free fermion BC}
\end{equation}
and functions $r^{\pm}(x,y)$, which are in the kernel of $\text{\ensuremath{\Delta^{(2)}}}$
subject to BC (\ref{eq: BC K(xx)} and \ref{eq: periodic BC K}). The
latter can be chosen in the form 

\begin{align*}
r^{+}(x,y) & =1,\\
r^{-}(x,y) & =1+\frac{2}{N}(x-y).
\end{align*}

To construct the solution for $g^{\pm}_{t}(x,y)$, we first find eigenfunctions
of the operator $\text{\ensuremath{\Delta^{(2)}}}$ subject to (\ref{eq:free fermion BC} and \ref{eq: periodic BC K}).
Let us look for the eigenvectors in the ``free fermion'' form 
\begin{equation}
\psi(x,y;z_{1},z_{2})=z^{x}_{1}z^{y}_{2}-z^{x}_{2}z^{y}_{1},\label{eq: free fermion wave function}
\end{equation}
which manifestly satisfies (\ref{eq:free fermion BC}). It is an eigenvector
of $\Delta^{(2)}$, corresponding to the eigenvalue 
\begin{equation}
\lambda_{2}(z_{1},z_{2})=\lambda_{1}(z_{1})+\lambda_{1}(z_{2}),\label{eq: lambda_2}
\end{equation}
where 
\begin{equation}
\lambda_{1}(z)=\frac{r}{z}+lz-1,\label{eq: lambda_1}
\end{equation}
and the values of $z_{1},z_{2}$ ensuring the (anti)periodic BC (\ref{eq: periodic BC K})
are solutions of 
\[
z^{N}_{i}=\mp1,i=1,2,
\]
i.e., they are the $N$th roots of $\mp1$ 
\[
z_{1,2}=e^{\mathrm{i}p^{\pm}_{j}},j=0,\dots,N-1,
\]
with 
\[
p^{\pm}_{j}=2\pi\frac{j}{N}+\frac{\pi}{2N}\pm\frac{\pi}{2N},\quad j=0,\dots,N-1.
\]

Thus, we find exactly $N(N-1)/2$ nonzero and inequivalent eigenvectors
\begin{equation}
\psi^{\pm}_{jk}(x,y)=\frac{1}{N}\left(e^{\mathrm{i}p^{\pm}_{j}x}e^{\mathrm{i}p^{\pm}_{k}y}-e^{\mathrm{i}p^{\pm}_{j}y}e^{\mathrm{i}p^{\pm}_{k}x}\right),0\leq j<k\leq N-1,\label{eq:basis}
\end{equation}
which form a complete basis in the complex linear space of complex-valued
functions on $\tilde{D}_{2}$ satisfying (\ref{eq:free fermion BC} and \ref{eq: periodic BC K}).
The basis obtained is orthonormal with respect to a semi-linear scalar
product 
\begin{equation}
\left\langle a,b\right\rangle =\sum_{1\leq x<y\leq N}\overline{a(x,y)}b(x,y),
\end{equation}
i.e.
\begin{equation}
\left\langle \psi^{\pm}_{j_{1}k_{1}},\psi^{\pm}_{j_{2}k_{2}}\right\rangle =\delta_{j_{1},j_{2}}\delta_{k_{1},k_{2}},
\end{equation}
where $0\leq j_{1}<k_{1}\leq N-1,\quad0\leq j_{2}<k_{2}\leq N-1.$
Having constructed this basis, we can immediately find the solution
for the function $g^{\pm}_{t}(x,y)$ subject to the initial conditions
$g^{\pm}_{0}(x,y)$
\begin{equation}
g^{\pm}_{t}(x,y)=\sum_{0\leq j<k\leq N-1}\psi^{\pm}_{jk}(x,y)e^{t\lambda_{2}\left(e^{\mathrm{i}p^{\pm}_{j}},e^{\mathrm{i}p^{\pm}_{k}}\right)}\left\langle \psi^{\pm}_{jk},g^{\pm}_{0}\right\rangle ,\label{eq: g_t(x_1.x_2)}
\end{equation}
where we should substitute
\begin{align}
g^{\pm}_{0}(x,y) & =K^{\pm}_{0}(x,y)-r^{\pm}(x,y).\label{eq: g_0^pm}
\end{align}
 The formulas (\ref{eq:g}, \ref{eq: g_t(x_1.x_2)}, and \ref{eq: g_0^pm})
give the final result for $K^{\pm}_{t}(x,y)$.

\bibliographystyle{plain}
\bibliography{cylinder}

@article{smoluchowski1916drei,
	title={Drei Vortr{\"a}ge {\"u}ber Diffusion, Brownsche Molekularbewegung und Koagulation von Kolloidteilchen [cz. 1]},
	author={Smoluchowski, Marian},
	year={1916},
	publisher={S. Hirzel}
}

@article{ovchinnikov1978role,
	title={Role of density fluctuations in bimolecular reaction kinetics},
	author={Ovchinnikov, AA and Zeldovich, Ya B},
	journal={Chemical Physics},
	volume={28},
	number={1-2},
	pages={215--218},
	year={1978},
	publisher={Elsevier}
}

@article{bramson1980asymptotics,
	title={Asymptotics for interacting particle systems on Z d},
	author={Bramson, Maury and Griffeath, David},
	journal={Zeitschrift f{\"u}r Wahrscheinlichkeitstheorie und verwandte Gebiete},
	volume={53},
	number={2},
	pages={183--196},
	year={1980},
	publisher={Springer}
}

@article{arratia1981limiting,
	title={Limiting point processes for rescalings of coalescing and annihilating random walks on Z d},
	author={Arratia, Richard},
	journal={The Annals of Probability},
	pages={909--936},
	year={1981},
	publisher={JSTOR}
}

@book{griffeath1979additive,
	title={Additive and cancellative interacting particle systems},
	author={Griffeath, David},
	year={1979},
	publisher={Springer}
}

@article{griffeath1978annihilating,
	title={Annihilating and coalescing random walks on Z d},
	author={Griffeath, David},
	journal={Zeitschrift f{\"u}r Wahrscheinlichkeitstheorie und Verwandte Gebiete},
	volume={46},
	number={1},
	pages={55--65},
	year={1978},
	publisher={Springer}
}

@article{erdos1974some,
	title={Some problems on random intervals and annihilating particles},
	author={Erd{\"o}s, P and Ney, Philip},
	journal={The Annals of Probability},
	pages={828--839},
	year={1974},
	publisher={JSTOR}
}

@article{masser2001correlation,
	title={Correlation functions for diffusion-limited annihilation, A+ A? 0},
	author={Masser, Thomas O and Ben-Avraham, Daniel},
	journal={Physical Review E},
	volume={64},
	number={6},
	pages={062101},
	year={2001},
	publisher={APS}
}

@article{ben1998complete,
	title={Complete exact solution of diffusion-limited coalescence, A+ A? A},
	author={Ben-Avraham, Daniel},
	journal={Physical review letters},
	volume={81},
	number={21},
	pages={4756},
	year={1998},
	publisher={APS}
}

@article{tribe2011pfaffian,
	author = {Roger Tribe and Oleg Zaboronski},
	title = {{Pfaffian Formulae for One Dimensional Coalescing and Annihilating Systems}},
	volume = {16},
	journal = {Electronic Journal of Probability},
	number = {none},
	publisher = {Institute of Mathematical Statistics and Bernoulli Society},
	pages = {2080 -- 2103},
	year = {2011},
	doi = {10.1214/EJP.v16-942},
	URL = {https://doi.org/10.1214/EJP.v16-942}
}

@article{tribe2012one,
	author = {Roger Tribe and Jonathan Yip and Oleg Zaboronski},
	title = {{One dimensional annihilating and coalescing particle systems as extended Pfaffian point processes}},
	volume = {17},
	journal = {Electronic Communications in Probability},
	number = {none},
	publisher = {Institute of Mathematical Statistics and Bernoulli Society},
	pages = {1 -- 7},
	year = {2012},
	doi = {10.1214/ECP.v17-2133},
	URL = {https://doi.org/10.1214/ECP.v17-2133}
}

@article{tribe2015erratum,
author = {Roger Tribe and Siu Yip and Oleg Zaboronski},
title = {{Erratum: one dimensional particle systems as extended Pfaffian point processes}},
volume = {20},
journal = {Electronic Communications in Probability},
number = {none},
publisher = {Institute of Mathematical Statistics and Bernoulli Society},
pages = {1 -- 2},
year = {2015},
doi = {10.1214/ECP.v20-4302},
URL = {https://doi.org/10.1214/ECP.v20-4302}
}

@article{garrod2018examples,
title={Examples of Interacting Particle Systems on Z as Pfaffian Point Processes: Annihilating and Coalescing Random Walks: B. Garrod et al.},
author={Garrod, Barnaby and Poplavskyi, Mihail and Tribe, Roger P and Zaboronski, Oleg V},
booktitle={Annales Henri Poincar{\'e}},
volume={19},
number={12},
pages={3635--3662},
year={2018},
organization={Springer}
}

@article{lushnikov1986binary,
	title={Binary reaction 1+ 1? 0 in one dimension},
	author={Lushnikov, AA},
	journal={Zh. Eksp. Teor. Fiz},
	volume={91},
	pages={1376--1386},
	year={1986}
}

@incollection{schutz2001exactly,
	title={Exactly solvable models for many-body systems far from equilibrium},
	author={Sch{\"u}tz, Gunter M},
	booktitle={Phase transitions and critical phenomena},
	volume={19},
	pages={1--251},
	year={2001},
	publisher={Elsevier}
}

@article{ben1995method,
	title={The method of inter-particle distribution functions for diffusion-reaction systems in one dimension},
	author={Ben-Avraham, Daniel},
	journal={Modern Physics Letters B},
	volume={9},
	number={15},
	pages={895--919},
	year={1995},
	publisher={World Scientific}
}

@article{glauber1963time,
	title={Time-dependent statistics of the Ising model},
	author={Glauber, Roy J},
	journal={Journal of mathematical physics},
	volume={4},
	number={2},
	pages={294--307},
	year={1963},
	publisher={American Institute of Physics}
}

@article{felderhof1971spin,
	title={Spin relaxation of the Ising chain},
	author={Felderhof, BU},
	journal={Reports on Mathematical Physics},
	volume={1},
	number={3},
	pages={215--234},
	year={1971},
	publisher={Elsevier}
}

@article{ginibre1965statistical,
	title={Statistical ensembles of complex, quaternion, and real matrices},
	author={Ginibre, Jean},
	journal={Journal of Mathematical Physics},
	volume={6},
	number={3},
	pages={440--449},
	year={1965},
	publisher={American Institute of Physics}
}

@article{borodin2009ginibre,
	title={The Ginibre ensemble of real random matrices and its scaling limits},
	author={Borodin, Alexei and Sinclair, Christopher D},
	journal={Communications in Mathematical Physics},
	volume={291},
	number={1},
	pages={177--224},
	year={2009},
	publisher={Springer}
}

@article{povolotsky2025representations,
	title={Representations of Hecke algebras and Markov dualities for interacting particle systems},
	author={Povolotsky, Alexander and Pyatov, Pavel and Tribe, Roger and Westbury, Bruce and Zaboronski, Oleg},
	journal={Annales de l'Institut Henri Poincare (B) Probabilites et statistiques},
	volume={61},
	number={2},
	pages={967--1020},
	year={2025},
	publisher={Institut Henri Poincar{\'e}}
}

@article{povolotsky2026coalescing,
	title={Coalescing and Annihilating particles on a torus.},
	author={Povolotsky, Alexander and Tribe, Roger and Zaboronski, Oleg},
	journal={to appear},
}

@article{torney1983diffusion,
	title={Diffusion-limited reactions in one dimension},
	author={Torney, David C and McConnell, Harden M},
	journal={The Journal of Physical Chemistry},
	volume={87},
	number={11},
	pages={1941--1951},
	year={1983},
	publisher={ACS Publications}
}

@article{balding1988diffusion,
	title={Diffusion-reaction in one dimension},
	author={Balding, David},
	journal={Journal of applied probability},
	volume={25},
	number={4},
	pages={733--743},
	year={1988},
	publisher={Cambridge University Press}
}

@article{balding1989diffusion1,
	title={Diffusion-controlled reactions in one dimension: Exact solutions and deterministic approximations},
	author={Balding, DJ and Green, NJB},
	journal={Physical Review A},
	volume={40},
	number={8},
	pages={4585},
	year={1989},
	publisher={APS}
}

@article{schutz1996nonequilibrium,
	title={Nonequilibrium correlation functions in the A+A? 0 system with driven diffusion},
	author={Sch{\"u}tz, GM},
	journal={Physical Review E},
	volume={53},
	number={2},
	pages={1475},
	year={1996},
	publisher={APS}
}

\end{document}